\documentclass[aps,pra,reprint,superscriptaddress,amsmath,amssymb]{revtex4-1}

\usepackage{color}
\usepackage{graphicx}
\usepackage{braket} 
\begin{document}

\title{Cavity driven Rabi oscillations between Rydberg states of atoms \\
  trapped on a superconducting atom chip}

\author{Manuel Kaiser}
\email{m.kaiser@uni-tuebingen.de}
\affiliation{Center for Quantum Science, Physikalisches Institut, 
Eberhard Karls Universit\"at T\"ubingen, Auf der Morgenstelle 14, 
D-72076 T\"ubingen, Germany}

\author{Conny Glaser}
\email{conny.glaser@uni-tuebingen.de}
\affiliation{Center for Quantum Science, Physikalisches Institut, 
Eberhard Karls Universit\"at T\"ubingen, Auf der Morgenstelle 14, 
D-72076 T\"ubingen, Germany}

\author{Li Yuan Ley}
\affiliation{Center for Quantum Science, Physikalisches Institut, 
Eberhard Karls Universit\"at T\"ubingen, Auf der Morgenstelle 14, 
D-72076 T\"ubingen, Germany}

\author{Jens Grimmel}
\affiliation{Center for Quantum Science, Physikalisches Institut, 
Eberhard Karls Universit\"at T\"ubingen, Auf der Morgenstelle 14, 
D-72076 T\"ubingen, Germany}

\author{Helge Hattermann}
\affiliation{Center for Quantum Science, Physikalisches Institut, 
Eberhard Karls Universit\"at T\"ubingen, Auf der Morgenstelle 14, 
D-72076 T\"ubingen, Germany}

\author{Daniel Bothner}
\affiliation{Center for Quantum Science, Physikalisches Institut, 
Eberhard Karls Universit\"at T\"ubingen, Auf der Morgenstelle 14, 
D-72076 T\"ubingen, Germany}

\author{Dieter Koelle}
\affiliation{Center for Quantum Science, Physikalisches Institut, 
Eberhard Karls Universit\"at T\"ubingen, Auf der Morgenstelle 14, 
D-72076 T\"ubingen, Germany}

\author{Reinhold Kleiner}
\affiliation{Center for Quantum Science, Physikalisches Institut, 
Eberhard Karls Universit\"at T\"ubingen, Auf der Morgenstelle 14, 
D-72076 T\"ubingen, Germany}

\author{David Petrosyan}
\affiliation{Center for Quantum Science, Physikalisches Institut, 
Eberhard Karls Universit\"at T\"ubingen, Auf der Morgenstelle 14, 
D-72076 T\"ubingen, Germany}
\affiliation{Institute of Electronic Structure and Laser, FORTH, 
GR-70013 Heraklion, Crete, Greece} 

\author{Andreas G\"unther}
\email{a.guenther@uni-tuebingen.de}
\affiliation{Center for Quantum Science, Physikalisches Institut, 
Eberhard Karls Universit\"at T\"ubingen, Auf der Morgenstelle 14, 
D-72076 T\"ubingen, Germany}

\author{J\'{o}zsef Fort\'{a}gh}
\email{fortagh@uni-tuebingen.de}
\affiliation{Center for Quantum Science, Physikalisches Institut, 
Eberhard Karls Universit\"at T\"ubingen, Auf der Morgenstelle 14, 
D-72076 T\"ubingen, Germany}

\begin{abstract}
Hybrid quantum systems involving cold atoms and microwave resonators can 
enable cavity-mediated infinite-range interactions between atomic spin systems
and realize atomic quantum memories and transducers for microwave to optical conversion. 
To achieve strong coupling of atoms to on-chip microwave resonators, 
it was suggested to use atomic Rydberg states with strong electric dipole transitions. 
Here we report on the realization of coherent coupling of a Rydberg transition 
of ultracold atoms trapped on an integrated superconducting atom chip 
to the microwave field of an on-chip coplanar waveguide resonator. 
We observe and characterize the cavity driven Rabi oscillations between 
a pair of Rydberg states of atoms in an inhomogeneous electric field near the chip surface. 
Our studies demonstrate the feasibility, but also reveal the challenges, of coherent
state manipulation of Rydberg atoms interacting with superconducting circuits. 
\end{abstract} 

\date{\today}

\maketitle

\section*{Introduction}
\label{sec:int}

Developing various hardware components for quantum information processing, storage and communication, 
as well as for quantum simulations of interacting many-body systems, is of fundamental importance for quantum 
science and technology. Quite generally, no single physical system is universally suitable for different tasks.
Hybrid quantum systems \cite{xiang2013,kurizki2015} composed of different subsystems with complementary 
functionalities may achieve high-fidelity operations by interfacing fast quantum gates \cite{Schoelkopf2008,Devoret2013} 
with long-lived quantum memories \cite{Lvovsky2009,Fleischhauer2005} and optical quantum communication 
channels \cite{Hammerer2010}. 

A particularly promising hybrid quantum system is based on an integrated superconducting atom chip 
containing a coplanar waveguide resonator and magnetic traps for cold neutral atoms. 
Microwave cavities can strongly couple with superconducting qubits \cite{Blais2004} and mediate 
quantum state transfer between the qubits and spin ensemble quantum memories \cite{Zhu2011,Kubo2011,Saito2013}. 
Cold atoms trapped near the surface of an atom chip possess good coherence properties 
\cite{sarkany2018faithful,Treutlein2004,Fortagh2007,Roux2008,Hermann-Avigliano2014} and 
strong optical (Raman) transitions, and are therefore suitable systems to realize 
quantum memories and optical interfaces \cite{Rabl2006,verdu2009strong,Hond2018,petrosyan2019microwave}.  
A promising approach for achieving strong coupling of atoms to on-chip microwave cavities is to employ 
appropriate Rydberg transitions with large electric dipole moments \cite{petrosyan2008quantum,petrosyan2009,Hogan2012}. 

Velocity-calibrated atoms prepared in circular Rydberg states and interacting 
one-by-one with a high-$Q$ microwave Fabry-Perot cavity (photon box) represent 
one of the most accurately controlled quantum optical systems \cite{haroche2013nobel}. 
Long-lived coherence of Rydberg state superpositions of atoms above a superconducting 
chip have been demonstrated \cite{hermann2014long}. 
Pioneering works have achieved electric-dipole coupling of Rydberg states of helium atoms 
in a supersonic beam with a coplanar microwave-guide \cite{Hogan2012} and, 
more recently, with a coplanar microwave resonator \cite{morgan2020coupling}.
So far, however, coupling of cold, trapped atoms with superconducting coplanar resonators 
have been achieved only for hyperfine magnetic-dipole atomic transitions \cite{hattermann2017coupling}. 

Here we demonstrate electric-dipole coupling of Rydberg states of ultra-cold atoms, 
trapped on a superconducting chip, to the microwave field of an on-chip coplanar microwave resonator. 
In doing so, we address the challenges associated with the inhomogeneous fields in the vicinity of the chip surface.
Our work is a stepping stone towards the realization of hybrid quantum systems with unique properties, 
such as switchable, resonator-mediated long-range interactions between the atoms \cite{petrosyan2008quantum}, 
conditional excitation of distant atoms and the realization of quantum gates 
\cite{petrosyan2008quantum, Pritchard2014, sarkany2015long, sarkany2018faithful}, 
coherent microwave to optical photon conversion via single atoms or atomic ensembles 
\cite{gard2017microwave,Covey2019microwave,petrosyan2019microwave}, and state transfer 
from solid state quantum circuits to optical photons via atoms \cite{tian2004interfacing}.

\section*{Results}
\label{sec:res}

%%%%%%%%%%%%%%%%%%%%%%%%%%%%%%%%%%%%%%%%%%%%%%%%%%%%%
\begin{figure*}[ht]
\includegraphics[width = 0.95\textwidth]{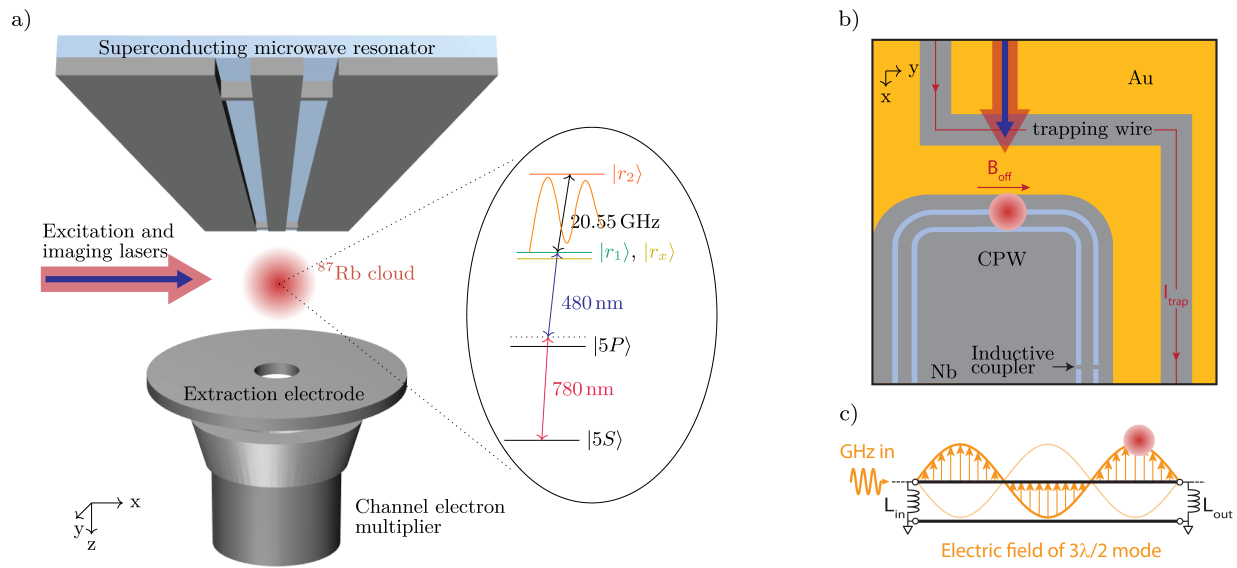}
\caption{\textbf{Schematics of the experimental system} (not to scale). 
\textbf{a} An ultracold cloud of $^{87}$Rb atoms is trapped at $130\,\mu$m distance from a CPW resonator. 
A pair of lasers with wavelengths $780\,$nm and $480\,$nm is used for two-photon excitation of the atoms 
from the ground state $\ket{5S}$ to a Rydberg state $\ket{r_1}$ (and $\ket{r_x}$) via an intermediate 
non-resonant state $\ket{5P}$. 
The MW coplanar waveguide cavity is inductively coupled via a feedline to an external MW source. 
In turn, the electric field of the cavity drives the atomic transition from state $\ket{r_1}$ 
to another Rydberg state $\ket{r_2}$, inducing damped Rabi oscillations. 
An extraction electrode below the cloud is used to apply a DC electric field for Stark-tuning the Rydberg 
transition frequency, for field-ionization of Rydberg atoms, and for extraction of the ions from the atom cloud. 
The ions are detected with a channel electron multiplier and adjacent detection electronics. 
\textbf{b} Bottom-view of the integrated superconducting atom chip containing a Z-shaped trapping wire and the CPW resonator: 
the trapping wire and the core of the MW resonator are made of superconducting niobium (gray); 
parts of the ground plane are coated with normal conducting gold (yellow); the bare sapphire chip surface is shown in blue. 
The magnetic trap with an offset field $B_{\mathrm{off}}$ at the center is generated by the current $I_{\mathrm{trap}}$ 
running through the trapping wire, persistent supercurrents around the cavity gaps, and external magnetic fields 
from macroscopic wires and coils \cite{bothner2013inductively}. 
\textbf{c} Circuit scheme of the inductively-coupled superconducting transmission line cavity and 
the corresponding standing-wave electric field of the third harmonic frequency $\omega_c = 2\pi\times 20.55$\,GHz. 
The atom cloud is trapped near one of the electric field antinodes.}
\label{fig:setup}
\end{figure*}
%%%%%%%%%%%%%%%%%%%%%%%%%%%%%%%%%%%%%%%%%%%%%%%%%%%%%

Our hybrid quantum system consists of a cloud of ultracold rubidium atoms trapped 
on an integrated superconducting (SC) atom chip containing a coplanar waveguide (CPW) resonator on the chip surface. 
Strong electric-dipole coupling between the atoms and the microwave (MW) cavity field requires, 
on the one hand, placing the atoms close to the chip surface, and, on the other hand, 
tuning the frequency of an appropriate Rydberg transition of the atoms into resonance 
with the fixed frequency of the cavity mode.  

\subsection*{Experimental apparatus}
	
The experimental system is illustrated in Fig.~\ref{fig:setup}\textbf{a}.  
The experiment is performed in an ultra-high vacuum chamber at a pressure of $6 \times 10^{-11}\,$mbar. 
The SC atom chip is attached to a He-flow cryostat with the surface temperature adjusted to 4.5\,K.
A chip-based magnetic trap for ultracold atoms is created via the field of the current in a Z-shaped 
trapping wire, the field associated with persistent SC currents in the resonator, and an external 
field generated by macroscopic wires and coils (see Fig.~\ref{fig:setup}\textbf{b}). 
The resulting trapping potential, with an offset field of $3.4\,$G at the trap center, has a harmonic shape
and can trap an atom cloud at a distance of $130\,\mu$m from the chip surface. 
A cloud of cold $^{87}$Rb atoms is loaded to the magnetic trap and shifted into the field mode 
of the standing-wave MW resonator, similar to Refs. \cite{hattermann2017coupling,bernon2013manipulation,bothner2013inductively}
(see the Methods section {M1} for more details).
The atom cloud is positioned close to an electric-field antinode of the third harmonic mode of 
the MW resonator at frequency $\omega_c \simeq 2 \pi \times 20.55$\,GHz (see Fig.\,\ref{fig:setup}\textbf{c}).  
The resonator is inductively coupled via a feedline to an external coherent and tunable MW source. 
A laser system is used for imaging the atomic cloud and for Rydberg excitation of the atoms. 
A DC electric field applied via the electrode can state-selectively ionize the Rydberg state atoms 
and extract the resulting ions for detection via a channel electron multiplier (CEM). 

\subsection*{Rydberg state excitation}

%%%%%%%%%%%%%%%%%%%%%%%%%%%%%%%%%%%%%%%%%%%%%%%%%%%%%
\begin{figure*}[ht]
\includegraphics[width = 0.95\textwidth]{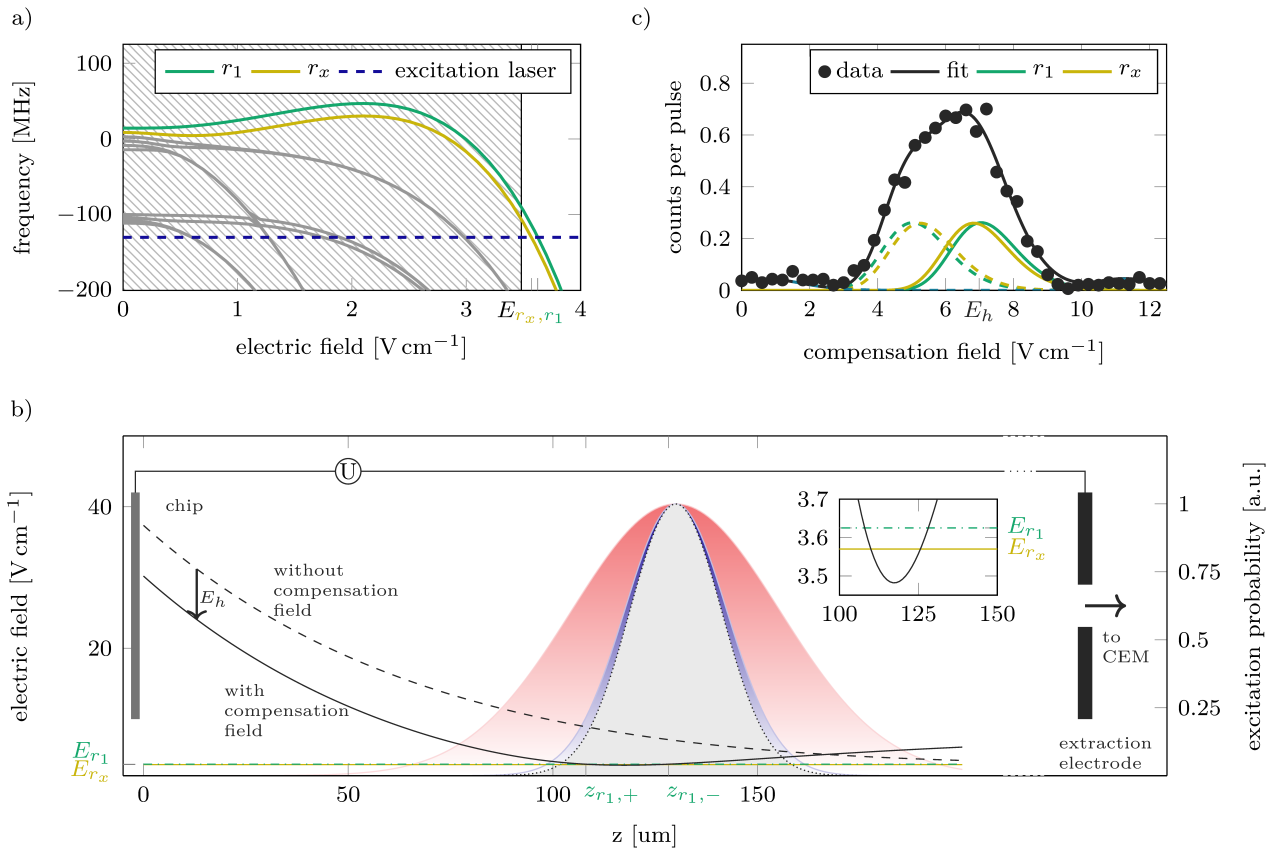}
\caption{\textbf{Laser excitation of Rydberg states of atoms in external fields.}
\textbf{a} Stark map of Rydberg states in the vicinity of the $48D$ state in a magnetic field of $3.4\,$G. 
The energies (frequency detuning relative to the $48D$ state in zero electric field) of the Stark states 
in the varying electric field are indicated by solid lines, while the dashed line indicates the frequency 
of the excitation laser detuned by $-2\pi\times 130\,$MHz. 
The target state $\ket{r_1}$ (green line) is resonantly excited by the laser in a field of $E_{r_1}=3.625\,$V/cm, 
and the nearby state $\ket{r_x}$ (yellow line) is resonant at $E_{r_x}=3.570 \,$V/cm.
Adsorbate field components parallel to the chip surface can not be compensated, 
preventing the excitation of Rydberg states in the hatched area. 
\textbf{b} Adsorbates on the chip surface produce an inhomogeneous electric field (black dashed line)
that falls off exponentially with the distance from the surface. 
A voltage $U$ applied between the chip and the extraction electrode creates a homogeneous electric field 
of $E_{\mathrm{h}} = 7.2 \,$V/cm that partially compensates the $z$-component of the adsorbate field 
at the position of the atom cloud (red profile). 
This leads to a minimum of the total field (black solid line) at a distance of $z_{\mathrm{min}}=117\,\mu$m 
from the chip surface (see the inset for a magnified view of this area). 
Due to non-compensated field components in the $xy$ plane, the total field 
has a parabolic form around the minimum (see Methods section {M2}). 
State $\ket{r_1}$ is resonantly excited in thin atomic layers at positions 
$z_{r_1}=108\,\mu$m and $128\,\mu$m where the field is $E_{r_1}$, while 
state $\ket{r_x}$ is resonantly excited at positions $z_{r_x}=110\,\mu$m and $126\, \mu$m where the field is $E_{r_x}$. 
The excitation probability at each of these positions is proportional to the product (gray profile)
of the cloud density (red profile) and the laser intensity (blue profile), and the corresponding 
atomic transition probabilities to states $\ket{r_1}$ and $\ket{r_x}$.
Applying a high-voltage ramp between the chip and the extraction electrode results in field 
ionization of Rydberg atoms and acceleration of the resulting ions towards the ion detector (CEM).
\textbf{c} Mean number of Rydberg excitations (ion counts, black dots) per laser pulse versus the applied compensation field. 
The experimental data is fitted by a sum of individual contributions from the $\ket{r_1}$ (green) and $\ket{r_x}$ (yellow) 
Rydberg states within two excitation layers (solid and dashed lines) each. 
From the fit we deduce the characteristic parameters of the adsorbate field.}
\label{fig:2}
\end{figure*}
%%%%%%%%%%%%%%%%%%%%%%%%%%%%%%%%%%%%%%%%%%%%%%%%%%%%%

The trapped atoms are excited from the ground state $\ket{5S}$ to a Rydberg state $\ket{r_1}$ (and to $\ket{r_x}$, see below) by a two-photon
transition, via an intermediate non-resonant state $\ket{5P}$, using a pair of laser pulses with wavelengths 
$780\,$nm and $480\,$nm (see Methods section {M1}). The atoms are subject to a
spatially varying DC (static) electric field, which is the sum of an inhomogeneous field produced by adsorbates
on the chip surface \cite{tauschinsky2010spatially, hattermann2012detrimental, chan2014adsorbate, sedlacek2016electric} 
and a controlled homogeneous compensation field produced by the extraction electrode (see Fig.~\ref{fig:setup}\textbf{a}). 
The electric field results in strong level shifts of the atomic Rydberg states. 
Figure \ref{fig:2}\textbf{a} shows the calculated Stark map of Rydberg states in the vicinity of the zero-field $48D$ state, 
taking into account the offset magnetic field of the trap. The calculation employs a diagonalization of the atomic
Hamiltonian in the presence of electric and magnetic fields \cite{zimmerman1979stark,grimmel2015measurement},
with the eigenvalues yielding the Rydberg energy spectrum for each field value. 

The electric field of the adsorbates falls off exponentially from the chip surface.  
An appropriate voltage applied to the extraction electrode creates a homogeneous compensation 
field between the chip surface and the electrode, which cancels the $z$ component of 
the adsorbate field at a desired position $z_{\mathrm{min}}$ within the atomic cloud. 
Since the adsorbate field is inhomogeneous, the total field strength increases in both directions 
from $z_{\mathrm{min}}$ along $z$. Furthermore, the adsorbates produce a non-vanishing field component 
parallel to the chip surface, in the $xy$ plane, which cannot be compensated for in our setup. 
This field component thus determines the minimal achievable 
field strength, while the total electric field in $z$ direction acquires a parabolic form
(see Fig. \ref{fig:2}\textbf{b} and the Methods section {M2}).
Hence, the resonance conditions for Rydberg excitations strongly vary across the atom cloud 
and each Rydberg state can be laser excited only in a thin ($\lesssim 0.3\, \mu$m) atomic layer. 
Since the energies of the Stark eigenstates depend only on the absolute value of the electric field, 
for each Rydberg state there are typically two resonant excitation layers, one on each side 
of the field minimum at $z_{\mathrm{min}}$ (see the inset of Fig.  \ref{fig:2}\textbf{b}). 
By varying the compensation field, these resonant layers can be shifted through the cloud 
with nearly constant layer separation. 

To match the position of resonant layers with that of the atomic cloud and the excitation laser beams, 
we vary the compensation field and count the number of Rydberg atoms excited by the laser with fixed 
detuning $-2\pi\times 130\,$MHz with respect to the zero-field $48D_{5/2}$ state. 
To this end, we prepare cold atomic clouds in the mode volume of the CPW cavity, 
at a distance of $130\, \mu$m from the chip surface.
Each of these clouds is exposed to a series of 300 excitation pulses of $1\,\mu$s duration, followed 
by Rydberg atom detection, at $3\,$kHz repetition rate, without significantly reducing the number of trapped atoms. 
After each pulse, the Rydberg excited atoms are ionized by a $1\,\mu$s 
electric field ramp and the resulting ions are subsequently detected by the CEM 
with a detection efficiency $>50$\% \cite{stibor2007calibration,guenther2009observing}. 
Figure \ref{fig:2}\textbf{c} shows the mean number of ion counts per excitation pulse 
as a function of the compensation field. We excite on average about one Rydberg atom per pulse, 
which allows us to disregard interactions between the Rydberg atoms. 
The resulting dependence of the ion count on the compensation field maps the atomic density 
distribution and the laser intensity profile in different excitation layers. The latter is deduced from fitting an appropriate model function 
to the experimental data (see Methods section {M2}). 
We then obtain that the exponential decay length of the adsorbate field from the chip surface 
is $\zeta  \simeq 70\,\mu$m, while the field component parallel to the chip surface has a value of $3.482\,$V/cm. 
For a compensation field of $7.2\,$V/cm, only two Rydberg states $\ket{r_1}$ and $\ket{r_x}$ 
are excited with significant probabilities in the center of the atom cloud. 
The calculated two-photon transition amplitudes between the ground $\ket{5S}$ and the Rydberg states 
$\ket{r_1}$ and $\ket{r_x}$ are approximately equal, which corresponds to similar excitation 
numbers $N_{r_1}$ and $N_{r_x}$. 

\subsection*{Tuning the Rydberg transition}

%%%%%%%%%%%%%%%%%%%%%%%%%%%%%%%%%%%%%%%%%%
\begin{figure*}
\includegraphics[width = 0.95\textwidth]{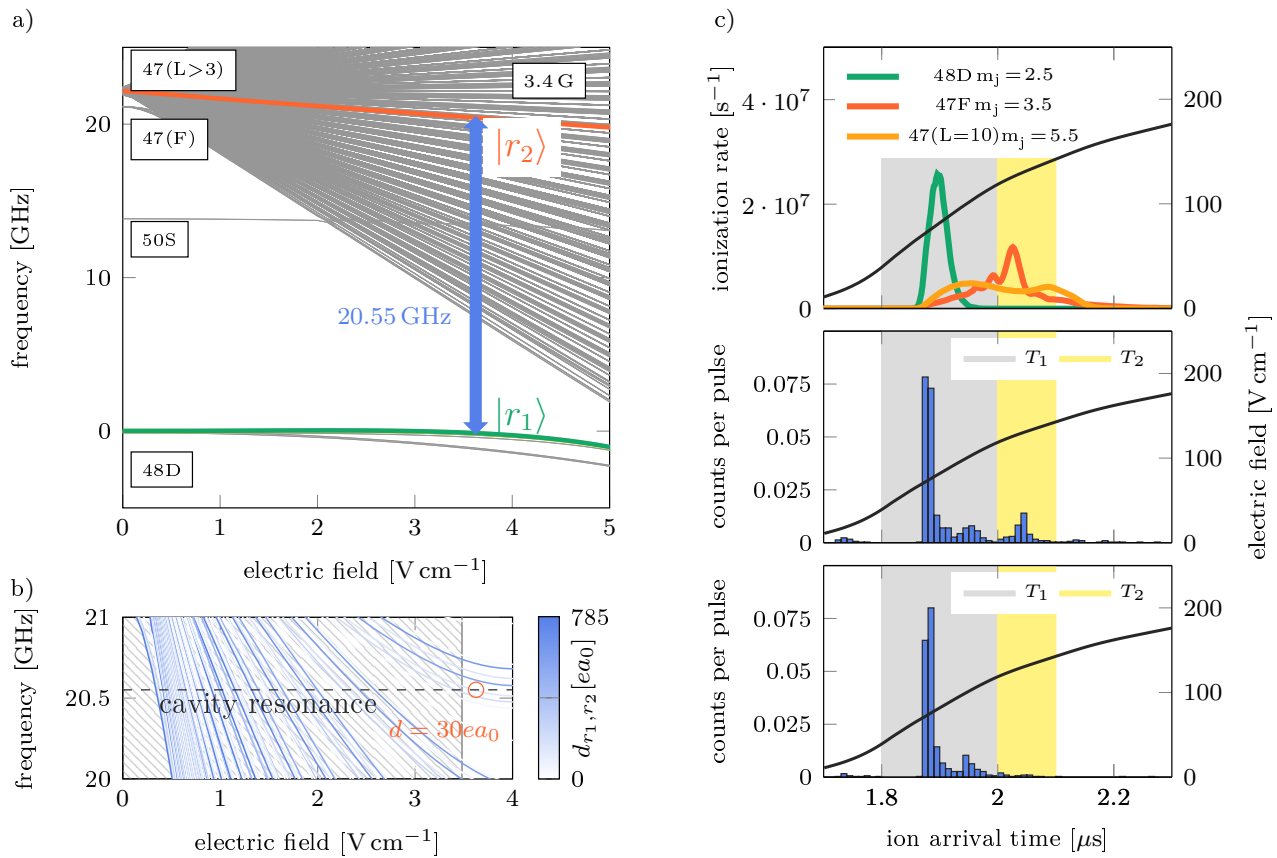}
\caption{\textbf{Rydberg transition and state detection.}
\textbf{a} Stark map of Rydberg states in the vicinity of $48D$ and $47(L\geq 3)$ states in a magnetic field of $3.4\,$G.
State $\ket{r_1}$ (green line) is resonantly excited by a laser in an electric field of $E_{r_1}=3.625\,$V/cm  
(see Fig. \ref{fig:2}\textbf{a}) and is coupled by a MW photon with frequency $\omega_{c} = 2 \pi \times 20.55\,$GHz 
to the Rydberg state $\ket{r_2}$ (red line) in the $47(L > 3)$ manifold.
\textbf{b} Frequencies (solid lines) and dipole moments (opacity) for transitions from state $\ket{r_1}$ to the states 
in the $n=47$ manifold, assuming in the calculations the total electric field along $x$, the magnetic field along $y$, 
and linear MW field polarization in the $z$ direction. The dashed horizontal line indicates the cavity resonance frequency 
$\omega_{c}$ and the transition to the resonant state $\ket{r_2}$ (red circle) has a dipole moment of $d = 30 ea_0$.  
\textbf{c} Upper panel: Examples of calculated ionization rates for three Rydberg states (green, red and orange lines)
upon the electric field ramp (black line).  
Central panel: Histogram of the ion arrival times (blue bars) as measured for a laser Rydberg excitation 
followed by a MW $\pi$-pulse of $250\,$ns duration and a subsequent electric field ramp (black solid line). 
The data is averaged over 25 experimental cycles with 300 excitation-interaction-detection sequences in each cycle. 
Defining two time intervals $T_1$ and $T_2$ permits distinguishing the ion signal from states $\ket{r_1}$ and $\ket{r_2}$.
Lower panel: Ion arrival time distribution without MW pulses.}
\label{fig:3}
\end{figure*}
%%%%%%%%%%%%%%%%%%%%%%%%%%%%%%%%%%%%%%%%%%

With an atom in Rydberg state $\ket{r_1}$, depending on the total electric and magnetic fields 
at the atomic position, there are many possible MW transitions to the Rydberg states in the $n=47$ manifold. 
In Fig. \ref{fig:3}\textbf{a} we show the calculated Stark map for the Rydberg states of atoms 
in the $3.4\,$G offset magnetic field of the trap and varying total electric field. 
Given the resonant frequency of the cavity field $\omega_c$ and the total electric field $E_{r_1}=3.625\,$V/cm
at the position of the resonantly laser-excited layer of atoms in state $\ket{r_1}$,
we deduce a suitable Rydberg transition to state $\ket{r_2}$ (see Fig. \ref{fig:3}\textbf{b}).
Our calculations also yield the dipole moments for different transitions (see Methods section {M3})
and we obtain the dipole moment $d \simeq 30 ea_0$ for the $\ket{r_1} \to \ket{r_2}$ transition, while the 
differential Stark shift between levels $\ket{r_1}$ and $\ket{r_2}$ is rather small, $-163\,$MHz/(V/cm).
Moreover, neighboring Rydberg states are sufficiently far off-resonant, which suppresses their excitation. 
 
\subsection*{Rydberg ionization signal}

Using the electric field ramp of the extraction electrode, we ionize the Rydberg state atoms and detect the resulting ions.
In principle, each state ionizes under the influence of an electric-field ramp with a characteristic time dependence,
which can be used to distinguish the Rydberg states \cite{gregoric2017quantum,gurtler2004}. 
In general, states with higher principal quantum numbers $n$ tend to ionize earlier, while states with higher 
orbital angular momenta $L$ and magnetic quantum numbers $m_j$ are ionized later \cite{gallagher2005rydberg}. In practice, the ionization 
signal from neighboring Rydberg states can have large temporal overlap, complicating their unambiguous discrimination.
In Fig.~\ref{fig:3}\textbf{c} upper panel, we show three examples of calculated ionization signals from different 
Rydberg states subject to the same electric field ramp (see Methods section {M4}). 
If we divide the ion arrival times into two intervals $T_1$ and $T_2$, then during $T_1$ we will detect all the
ions from state $48D$ and some of the ions from states $n=47$, while the ions detected during $T_2$ will only
be from states $n=47$.   

In the experiment, we populate via laser excitation states $\ket{r_x}$ and $\ket{r_1}$ 
with comparable probabilities,  $N_{r_x}/N_{r_1} = a \approx 1$, and then 
couple $\ket{r_1}$ to state $\ket{r_2}$ via the resonant MW cavity field. 
The electric field ramp results in the Rydberg state ionization and detection of $N = N_{r_x} + N_{r_1} + N_{r_2}$ ions. 
All the ions $N_{r_x}$ and $N_{r_1}$ from states $\ket{r_x}$ and $\ket{r_1}$ are detected 
during the time interval $T_1 = 1.8-2.0\,\mu$s, while we estimate that the $N_{r_2}$ ions from state $\ket{r_2}$ 
are detected during the time interval $T_2 = 2.0-2.1\,\mu$s with probability $p=0.34$ 
and during $T_1$ with probability $(1-p)=0.66$. 
Hence, the number of ions $N_{T_1}$ and $N_{T_2}$ detected during $T_1$ and $T_2$ are
$N_{T_1} = N_{r_x} + N_{r_1} + (1-p)N_{r_2} \simeq (1+a) ( N_{r_1} + N_{r_2}) - pN_{r_2}$ and $N_{T_2} = p N_{r_2}$, 
while $N_{T_1} + N_{T_2} = N$. 
We then obtain that the population of state $\ket{r_2}$ is 
$\rho_{22} \equiv \frac{N_{r_2}}{N_{r_1}+N_{r_2}}=\frac{(1+a)}{p} \frac{N_{T_2}}{N} = \frac{1+a}{p} p_2$, where $p_2 \equiv N_{T_2}/N$.

In Fig.~\ref{fig:3}\textbf{c} middle panel, we show the measured ion signal after applying a MW $\pi$-pulse, 
which would ideally transfer all of the population of state $\ket{r_1}$ to state $\ket{r_2}$. The ion arrival times
have a large peak in $T_1$ originating from all states $\ket{r_x}$, $\ket{r_1}$, $\ket{r_2}$, and a smaller peak
in $T_2$ that we attribute only to state $\ket{r_2}$. With $N_{T_1}/N_{T_2} \approx 5.6$, the population of state
$\ket{r_2}$ is $\rho_{22} \simeq 0.9$, consistent with a weakly damped half of a resonant Rabi cycle.
Without the MW pulse, all the population remains in states $\ket{r_x}$ and $\ket{r_1}$ and no peak is observed 
in $T_2$, as shown in the lower panel of Fig.~\ref{fig:3}\textbf{c}. 
	
\subsection*{Cavity driven Rydberg transition}

%%%%%%%%%%%%%%%%%%%%%%
\begin{figure*}
\includegraphics[width = 0.95\textwidth]{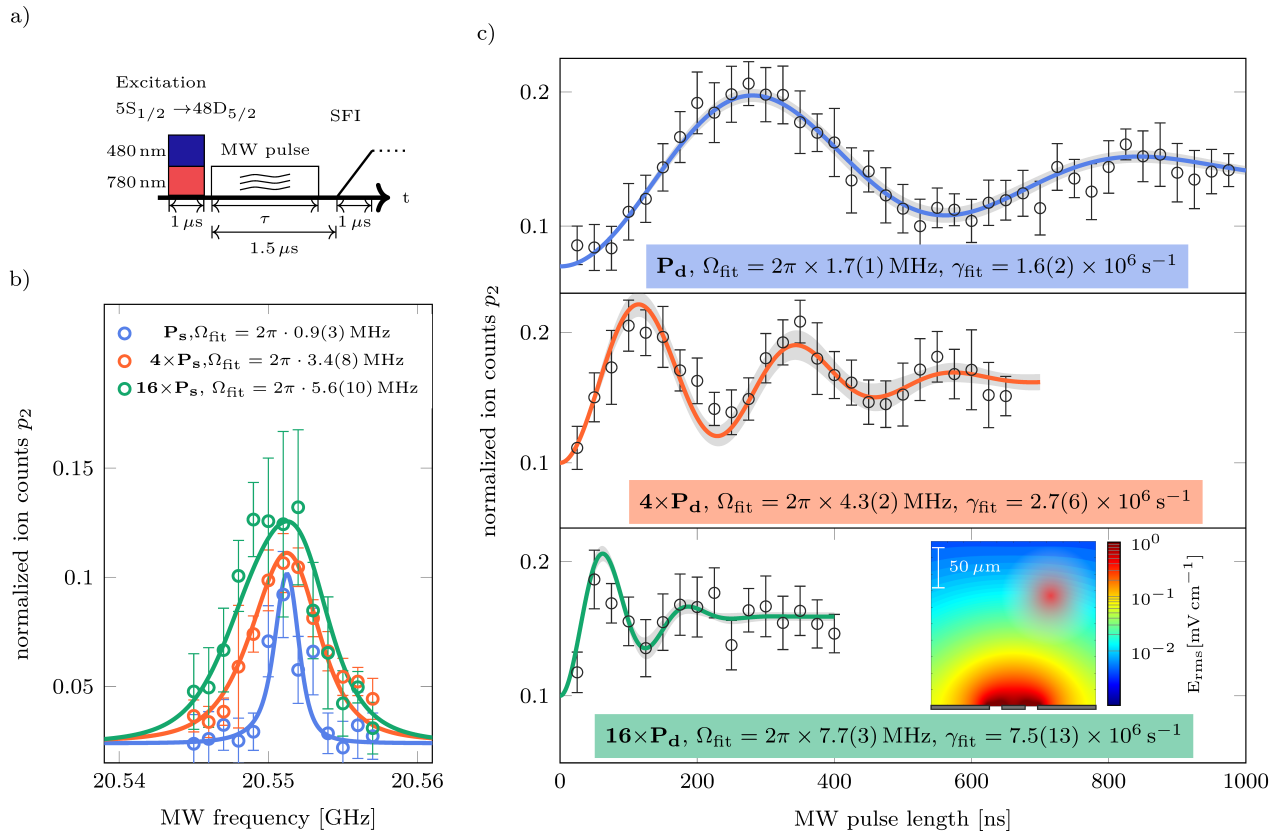}
\caption{\textbf{Coupling Rydberg transition to pumped MW cavity mode.} 
  \textbf{a} Schematics of the experimental sequence involving optical excitation of the atoms to the Rydberg state 
  by a two-photon laser pulse of $1\,\mu$s duration, injection of a MW pulse of variable power and duration 
  $\tau \leq 1.5 \,\mu$s into the cavity, interaction of the Rydberg excited atoms with the MW cavity field, 
  and selective field ionization (SFI) and detection of the Rydberg atoms by a $1\,\mu$s electric field ramp 
  from $7.2\,$V/cm to about $210\,$V/cm. 
  In each experimental cycle, the atom cloud is exposed to a series of $300$ excitation-interaction-detection 
  sequences at a repetition rate of $3\,$kHz.
  MW leakage through the electronic switches and background counts lead to a small offset of the ion signal.
  \textbf{b} Normalized ion count $p_2 = N_{T_2}/N$ (open circles), proportional to the population $\rho_{22}$ of Rydberg state $\ket{r_2}$,
  versus the frequency of the injected into the cavity MW pulses of duration $\tau = 1.5\,\mu$s,
  for three different powers $P_s,4P_s,16P_s$ (blue, red, green) of the MW source.    
  Each data point is obtained by averaging over 10 experimental cycles, with error bars denoting $\pm2$ standard deviations. 
  We fit the data by theoretical excitation spectra (solid lines of the same color) to deduce the Rabi frequencies 
  $\bar{\Omega}$ of the MW field resonant with the $\ket{r_1} \to \ket{r_2}$ transition.  
  \textbf{c} Dynamics of population of Rydberg state $\ket{r_2}$ (proportional to ion count $p_2$) for three different 
  powers $P_d,4P_d,16P_d$ of the injected MW field at frequency $\omega_{\mathrm{MW}} = 20.551\,$GHz. 
  Each data point (open circle) is obtained by averaging over 25 experimental cycles, with error bars denoting $\pm2$ standard deviations.
  Rabi frequencies $\Omega_{\mathrm{fit}}$ and decoherence rates $\gamma_{\mathrm{fit}}$ are obtained by fitting  the model function 
  (see the text and Methods section {M5} for details), with the fitted curves (solid lines) shown with 95\% confidence bounds (gray).  
  The inset shows the MW cavity field strength in the $xz$ plane for a rms ground state voltage of $3\,\mu$V (see Methods section {M3}), 
  with the cloud placed at a distance of $z \simeq 130\,\mu$m from the chip surface (bottom) and laterally displaced from the central 
  cavity conductor by $x \simeq 50\,\mu$m. The gray bars indicate the position of the CPW.}
\label{fig:4}
\end{figure*}
%%%%%%%%%%%%%%%%%%%%%%%%

Having described all the necessary ingredients of our system, we now turn to the demonstration of 
coherent coupling of the atomic Rydberg transition to an externally pumped MW field mode of the cavity. 
The experimental sequence is illustrated in Fig.~\ref{fig:4}\textbf{a}: after the two-photon laser excitation of 
the atoms to the Rydberg state $\ket{r_1}$, we inject into the cavity a MW pulse of variable power and duration. 
The Rydberg atom interacts with the cavity field on the transition $\ket{r_1}-\ket{r_2}$ for up to $1.5\, \mu$s. 
The cavity induced population transfer between the Rydberg states is then detected via selective field ionization 
and time-resolved ion counting. 

The interaction of the atom with the MW cavity field can be well approximated by a two-level model. 
The field mode of the cavity with frequency $\omega_c = 2\pi\times 20.55\,$GHz is resonant with 
the atomic transition $\ket{r_1}-\ket{r_2}$, while the next nearest Rydberg transition is detuned by 
at least $15\,$MHz (see Fig. \ref{fig:3}\textbf{b}), which exceeds the cavity linewidth $\kappa = 2 \pi \times 9\,$MHz
and the maximal achieved Rabi frequency $\Omega_{\mathrm{max}} \simeq  2 \pi \times 8\,$MHz.
There are no near resonant transitions from state $\ket{r_x}$. Decay to other states is negligible
during the entire excitation, interaction and measurement time of $\sim 3\,\mu$s, which is much shorter than 
the Rydberg state lifetime $1/\Gamma \approx 60 \, \mu$s \cite{mack2015all,beterov2009quasiclassical}.  

In Fig.~\ref{fig:4}\textbf{b} we show the Rydberg transition spectrum of the atoms interacting with 
the intracavity MW field pumped by the external source of variable frequency at three different powers.
The exposure time of $1.5\,\mu$s is sufficiently long for the atomic populations to attain the steady state.
We fit the observed spectrum with a model function corresponding to the steady-state population $\rho_{22}$
of the Rydberg state $\ket{r_2}$, while the Rabi frequency of the MW field also depends on the frequency of the injected into the cavity field (see Methods section {M5}). We assume that the main contribution 
to the atomic spectral linewidth comes from the inhomogeneous broadening $\Delta\omega_{12} \simeq 2\pi \times 1.1\,$MHz 
of the transition $\ket{r_1}-\ket{r_2}$ due to the differential Stark shift of $-163\,$MHz/(V/cm) between 
levels $\ket{r_1}$ and $\ket{r_2}$ in the inhomogeneous electric field varying by $0.0067\,$V/cm within
the resonantly excited atomic layer of widths $\lesssim 0.3\, \mu$m (see Methods section {M3}). 
In comparison, the natural linewidth (decay rate) $\Gamma \simeq 2\pi\times 2.7\,$kHz of the Rydberg 
state is negligible. For stronger pumping fields, the spectrum is dominated by power broadening. 

For each pumping power, we extract the peak Rabi frequency $\bar{\Omega}$ at the pump MW frequency 
$\omega_{\mathrm{MW}} \simeq \bar{\omega}_{12}$ resonant with the mean transition frequency of the atoms
in the resonant layer. We then obtain that the peak Rabi frequency $\bar{\Omega}$ scales approximately 
as square root of the pumping power $P$, as expected for a resonant one-photon transition. 

By changing the detuning of the excitation lasers, we can excite the Rydberg states of atoms in 
a different resonant layer. This in turn will shift the atomic Rydberg transition out of the cavity resonance, 
causing the signal in Fig.~\ref{fig:4}\textbf{b} to disappear. Using a MW antenna from outside 
the chamber, we can recover the ionization signal for an appropriate MW frequency, which verifies that 
the Rydberg MW transition is indeed driven by the cavity field. 
The cavity and the atomic resonance are, however, not perfectly aligned, which leads to a small asymmetry 
of the resonance profiles that becomes more pronounced for increasing MW power, 
as seen in Fig.~\ref{fig:4}\textbf{b}. 

We next fix the frequency of the injected MW field to $\omega_{\mathrm{MW}} = 2\pi\times 20.551\,$GHz, and investigate the 
dynamics of the Rydberg transition by varying the duration of the MW pulse for three different input powers. 
In Fig.~\ref{fig:4}\textbf{c} we show the normalized ion count $p_2$ from the upper Rydberg state
$\ket{r_2}$ as a function of the MW pulse duration. We observe damped Rabi oscillations between states
$\ket{r_1}$ and $\ket{r_2}$ with the Rabi frequencies that scale approximately as $\Omega \propto \sqrt{P}$,
similarly to the steady-state case above. We approximate the population difference between the Rydberg states as 
$\bar{D}=\rho_{11} -\rho_{22} = \exp (- t^2/\tau_{\mathrm{damp}}^2 )  \cos (\bar{\Omega} t)$, which follows from a simple model 
(see Methods section {M5}) that assumes two-level atoms coherently driven by spatially varying Rabi frequencies 
corresponding to an inhomogeneous MW cavity field distribution across the laterally displaced atomic ensemble, 
as shown in the inset of Fig.~\ref{fig:4}\textbf{c}.
The corresponding damping rate $\gamma = 1/\tau_{\mathrm{damp}}$ then scales approximately linearly with the Rabi frequency,  
$\gamma \simeq \frac{b}{\sqrt{2}} \bar{\Omega} $ with the proportionality constant $b \approx 0.2$ that quantifies the
relative change of cavity MW field across the atomic cloud along the $x$ direction.

\section*{Discussion}
\label{sec:dis}

We have demonstrated coherent coupling of ultracold Rb atoms, trapped on an integrated atom chip and 
laser excited to the Rydberg states, to a superconducting MW resonator. We used DC electric fields to fine-tune the atomic Rydberg
transition and observed resonant Rabi oscillations between a pair of Rydberg states driven by the MW field of the cavity
pumped by an external coherent source. The observed damping of the Rabi oscillations in our system is dominated by the
spread of Rabi frequencies for the Rydberg atoms in different positions of the resonantly excited by the laser layer.
In comparison, the real single atom decoherence is much smaller, and individual Rydberg atoms are expected to exhibit 
much longer coherence times. 

We note, that the damping rate of the Rabi oscillations originating from the variations of the cavity field at different
positions within the atom cloud is significantly reduced by the resonant laser excitation of Rydberg states of atoms in a thin 
layer at a well defined distance from the atom chip. Better positioning of the atomic cloud, reducing the lateral size of the excitation layer, and using narrow-line excitation
lasers can further reduce or completely eliminate the cavity field variations for the Rydberg excited atoms, leading
to much longer coherence times of the MW Rydberg transition of the atoms.

The atoms in our system are trapped relatively far from the CPW cavity, $130\,\mu$m from the chip surface, and the 
resulting vacuum Rabi frequency  $\Omega_{\mathrm{vac}} = d E_c^{(0)}/\hbar = 2\pi\times 0.6\,$kHz is much smaller than
the cavity linewidth $\kappa \simeq 2\pi\times 9\,$MHz. Several improvements can be made: The atoms can be brought
much closer to the chip surface and, e.g., at a distance of $\sim 10\,\mu$m the cavity field strength $E_c^{(0)}$ 
will increase by more than an order of magnitude. At smaller distances, the stronger adsorbate fields should be treated 
more carefully and a pair of Rydberg-Stark eigenstates with much larger dipole moments $d \simeq 10^3 ea_0$ and small
differential Stark shift should be found and used.  Furthermore, there has been great progress in increasing the 
quality factors of superconducting coplanar waveguide resonators, already exceeding $10^6$ at $\sim 50\,$mK temperatures
\cite{megrant2012planar}, which also enables coherent operations of SC qubits. 
This will then bring the strong-coupling regime of single Rydberg atoms and SC qubits 
to single MW cavity photons within the experimental reach.

\newpage

\section*{Methods}
\label{sec:met}

\subsection*{M1 EXPERIMENTAL SYSTEM}
\label{met:exp_system}
\textbf{Atomic cloud.} 
Starting with $^{87}$Rb atoms in a conventional magneto-optical trap, about $5 \times 10^7$ atoms 
in the ground state $\ket{5S_{1/2}, F=2, m_F=2}$ are loaded into a magnetic quadrupole trap. 
The atoms are then moved to a Ioffe trap for evaporative cooling. 
Using optical tweezers, the ultracold atomic cloud is transferred to a superconducting atom chip, 
which is mounted on the cold-finger of a He-flow cryostat. 
At the chip, the spin-polarized atomic cloud is loaded into a magnetic microtrap  
generated by a current in the Z-shaped trapping wire and other external homogeneous fields. 
Additional bias fields are used to finally position the cloud close to a coplanar waveguide resonator. 
Using standard absorption imaging, we verify that the cloud is trapped at $130\,\mu$m distance 
to the chip surface.
The trapping potential has a parabolic shape with trap frequencies $\omega_{x,y,z} = 2 \pi \times 61$\,Hz.
The offset field at the trap center is about $3.4\,$G. 
The atom cloud has spherical shape with a cloud diameter (full width at half maximum) of $60\,\mu$m.
The atom number in the cloud is $7 \times 10^4$, the peak density is $2.7 \times 10^{11}\, \mathrm{cm}^{-3}$
and the temperature is $1 \, \mu$K. 
More details on the loading and trapping procedure are given in Ref.\,\cite{hattermann2017coupling}.\\

\textbf{Rydberg lasers.} 
The atoms are excited from the ground state $\ket{5S_{1/2}}$ to the Rydberg states 
by a pair of laser pulses with wavelengths $780\,$nm and $480\,$nm and duration of $1\,\mu$s, 
while the intermediate state $\ket{5P_{3/2}}$ is detuned by $160\,$MHz. 
Both laser beams are centered at the atomic cloud. 
With beam waists ($1/e^2$-radii) of $5.45\,$mm and $0.025\,$mm and laser powers of $0.5\,$mW and $25\,$mW, 
the peak intensities of the red and blue lasers are $0.001\,$W/cm$^2$ and $2500\,$W/cm$^2$, respectively.
The red laser is phase locked to a commercial frequency comb with an absolute frequency accuracy of $2\,$kHz and linewidth below $10\,$kHz. 
The blue laser is stabilized with a wavelength meter (HighFinesse WSU) to an absolute frequency accuracy 
better than $2\,$MHz and a linewidth better than $200\,$kHz.
In zero field conditions, the single-photon Rabi frequencies for individual transitions
$5S_{1/2}-5P_{3/2}$ and $5P_{3/2}-48D_{5/2}$, averaged over all magnetic sub-states and possible transitions, 
are $2\pi\times 3.6\,$MHz and $2\pi\times 36\,$MHz, respectively. 
The resulting effective two-photon Rabi-frequency for the Rydberg excitation $5S_{1/2}-48D_{5/2}$ is 
$2\pi\times 400\,$kHz.
Taking into account the external fields and applying the selection rules for appropriate laser polarization, 
the two-photon Rabi-frequency for the transition $\ket{5S_{1/2}, F=2, m_F=2} \to \ket{r_1}$ and $\ket{r_x}$
is reduced to about $\Omega_{gr}=2\pi\times 30\,$kHz;
the ratio $a$ between these Rabi frequencies depends on the unknown direction of the lateral surface field 
(see Sec.~{M2} below) and is assumed $a \approx 1$.
With this Rabi frequency and a pulse duration of $1\,\mu$s applied to the ensemble of $7 \times 10^4$ ground state atoms, 
we excite on average about one atom to the Rydberg state $\ket{r_1}$. \\

\textbf{Microwave cavity.} 
The on-chip superconducting coplanar waveguide cavity is a $\lambda/2$ transmission line resonator 
with the fundamental mode frequency $\omega_0=2 \pi \times 6.85\,$GHz. 
To drive the transition between the Rydberg states $\ket{r_1}$ and $\ket{r_2}$, 
we use the third harmonic mode of the resonator having the frequency $\omega_c = 2\pi\times 20.55\,$GHz, 
the linewidth $\kappa = 2 \pi \times 9\,$MHz and the corresponding quality factor $Q \approx 2300$.
Within the temperature range where the resonator is superconducting, the resonance frequency 
can be tuned by about $30\,$MHz.
For driving purposes, the resonator is inductively coupled to a MW feedline and connected 
to a commercial MW signal generator (Rohde \& Schwarz SMF100A) with a linewidth $<1\,$Hz.
The MW electric field in the resonator is then
\begin{equation}
E_c \propto \frac{\sqrt{P}}{\kappa/2 - i(\omega_{\mathrm{MW}}  - \omega_c)} , \label{eq:EcvsMW}
\end{equation}
where $P$ is the power of the injected into the resonator MW field at frequency $\omega_{\mathrm{MW}}$.

\subsection*{M2 RYDBERG EXCITATION}
\label{met:ryd_excitation}
\textbf{Surface fields.} 
The adsorbates on the chip surface produce an inhomogeneous electric field. 
We approximate the $z$ component of this adsorbate field perpendicular to the chip surface 
as $E_{\mathrm{ad},z}(z) = E_{0} e^{-z/\zeta}$, where $\zeta$ is a decay length. 
This component is partially compensated by a homogeneous field $E_{\mathrm{h}}$
between the extraction electrode and the grounded atom chip surface. 
The compensation field can be freely adjusted by the voltage $U$ applied 
to the extraction electrode, $E_{\mathrm{h}}/U = 3\,$(V/cm)/V. 
The inhomogeneous adsorbate distribution leads to an additional field component 
parallel to the chip surface, $E_{\mathrm{ad},xy}$, which can not be compensated for in our system. 
The total electric field  
\begin{equation}
E(z) = \sqrt{(E_0 e^{-z/\zeta} - E_{\mathrm{h}})^2 + E_{\mathrm{ad},xy}^2}
\end{equation} 
can then be tuned via $E_{\mathrm{h}}$ and has a parabolic form around the field minimum 
$E(z_{\mathrm{min}})=E_{\mathrm{ad},xy}$ at $z_{\mathrm{min}} = \zeta \ln \frac{E_0}{E_{\mathrm{h}}}$. \\

\textbf{Excitation layers.}
The Stark shifts of the atomic Rydberg levels in a DC electric field depend only on the absolute field strength. 
Within the atomic cloud, there can be two layers where the electric field has a specific value 
$E_r >E(z_{\mathrm{min}}) = E_{\mathrm{ad},xy}$. These layers are positioned on both sides of the field minimum at
\begin{equation}
z_{r,\pm} = z_{\mathrm{min}} - \zeta \ln {\left(1 \pm \sqrt{\frac{E_r^2 - E_{\mathrm{ad},xy}^2}{E_{\mathrm{h}}^2}}\right)} .          
\end{equation}
A Rydberg state $\ket{r}$ that is resonant at a specific field strength $E_r$ 
is thus excited in two excitation layers at position $z_{r,\pm}$. 
For a field $E_r = E_{\mathrm{ad},xy}$ both layers merge at $z_{r,\pm}=z_{\mathrm{min}}$, 
while no excitation can take place for states that are resonant in fields $E_r < E_{\mathrm{ad},xy}$. 

The local field gradient in a resonant excitation layer is 
\begin{equation}
\alpha_{r,\pm} = \left.\frac{dE}{dz}\right|_{z_{r,\pm}} = \frac{1}{\zeta} \frac{E_{\mathrm{h}}^2}{E_r} (\beta \mp \sqrt{\beta_r} )
\label{eq:Ergrad}
\end{equation}
with $\beta_r = \sqrt{E_r^2 - E_{\mathrm{ad},xy}^2}/E_{\mathrm{h}}$. 
Assuming the two-photon Rabi frequency $\Omega_{gr}$ is small compared to the linewidth of the excitation laser $\Delta \omega_{\mathrm{las}}$,
the width of the excitation layer can be estimated as 
\begin{equation}
\Delta z_{r,\pm} = \frac{\Delta\omega_{\mathrm{las}}}{d_r \alpha_{r,\pm}}, \label{eq:Deltaz}
\end{equation}
where $d_r$ is the Stark gradient, or static dipole moment, of state $\ket{r}$. \\

\textbf{Excitation probability}
With the excitation layers oriented parallel to the chip surface, 
the Rydberg excitation probability of the atoms within a single layer at $z$ 
is proportional to the atomic line density and the laser intensity profile,  
\begin{eqnarray}
n(z) & \propto & \exp{ \left( -\frac{(z-z_{\mathrm{cloud}})^2}{2\sigma^2} \right) } , \label{eq:atomicdensity} \\
I(z) & \propto & \exp{ \left( -\frac{2(z-z_{\mathrm{beam}})^2}{w^2} \right) } , \label{eq:laserprofile}
\end{eqnarray}
where $z_{\mathrm{cloud}}$ is the position of the atomic cloud, $\sigma$ is the cloud radius,
$z_{\mathrm{beam}}$ is the laser beam center and $w$ is the beam waist. 
For a two-photon transition, $\Omega_{gr} \propto \sqrt{I_{780} I_{480}}$, 
the effective beam waist is $w = w_{780} w_{480} / \sqrt{w_{780}^2 + w_{480}^2}$ 
with the individual beam waists $w_{780,480}$ assumed to overlap at the same position. 
The total Rydberg excitation probability is given by the sum over all possible 
transitions in the corresponding excitation layers:
\begin{equation}
P_{\mathrm{Ry}} =\sum_{r}\sum_{s=\pm} A_{r,s}\, n(z_{r,s})\, I(z_{r,s})\label{eq:prob}
\end{equation}
where the strength coefficients $A_{r,\pm}$ depend on the dipole matrix element 
of the corresponding transition. \\

\textbf{Fit to the data.} 
We use Eq.~(\ref{eq:prob}) to fit the experimental data in Fig.~\ref{fig:2}\textbf{c}
taking into account all states and their corresponding excitation layers. 
For each state $\ket{r}$, we determine the resonance field value $E_r$ 
from the Rydberg Stark map calculations in Fig.~\ref{fig:2}\textbf{a}
with the excitation laser detuning set to $-130\,$MHz with respect to the zero-field $48D_{5/2}$ state. 
Atomic cloud and beam parameters are $z_{\mathrm{cloud}} = z_{\mathrm{beam}} = 130\,\mu$m, $\sigma = 25\,\mu$m 
and $w \approx w_{480} = 25\, \mu$m, leaving only the strength coefficients $A_r$ 
and the adsorbate field parameters $E_0$, $\zeta$ and $E_{\mathrm{ad},xy}$ as free fitting parameters. 
We then obtain $E_{0} = 37.2 \,$V/cm, $\zeta = 70\, \mu$m and $E_{\mathrm{ad}, xy} = 3.482\,$V/cm. 
Rydberg excitations mostly occur in the resonant excitation layers of the two states $\ket{r_1}$ and $\ket{r_x}$, 
as other states can be resonantly excited in different electric fields, at positions outside of the atomic cloud.

For a compensation field of $E_{\mathrm{h}} = 7.2 \,$V/cm, the two excitation layers for state $\ket{r_1}$ 
(resonant with the excitation laser at total field $E_{r_1}= 3.625 \,$V/cm) are located at 
$z_{r_1,+} = 108\,\mu$m and $z_{r_1,-} =128\,\mu$m. 
With the two-photon excitation linewidth $\Delta \omega_{\mathrm{las}} = 2\pi\times 2\,$MHz, 
the Stark gradient $d_{r_1}= -300\,$MHz/(V/cm) and the field gradients $\alpha_{r_1,\pm}$,
the corresponding layer widths are $\Delta z_{r_1,+}=0.21\,\mu$m and $\Delta z_{r_1,-}=0.28\,\mu$m. 
Since the atomic cloud is centered at $z = 130\,\mu$m, dominant excitation takes place in the $z_{r_1,-}$ layer. 
Energetically close states are excited in layers at different distances from the chip. 
They contribute to the ion signal, but do not participate in the MW transition between the Rydberg states 
and therefore only affect the contrast in the Rabi oscillation measurements.

\subsection*{M3 MICROWAVE RYDBERG TRANSITION}
\label{met:mw_coupling}

\textbf{Matrix elements.} 
The energy spectrum of Rydberg states in external fields can be calculated by diagonalization 
of the atomic Hamiltonian, including the interactions with the magnetic and electric fields, 
in the basis of zero-field states $\ket{n,l,j,m_j}$ \cite{zimmerman1979stark}. 
We compute the energy eigenvalues and the corresponding eigenvectors 
\begin{equation}
\ket{r} = \sum_{n,l,j,m_j} \beta^{n,l,j,m_j} \ket{n,l,j,m_j} ,
\label{eq:epsr}
\end{equation}
where $\beta^{n,l,j,m_j}$ are the normalized amplitudes of the basis states. 
The dipole moment for the transition between a pair of Rydberg states $\ket{r_i}$ 
and $\ket{r_f}$ is then \cite{grimmel2015measurement}
\begin{equation}
\vec{d} = e \braket{r_f| \vec{r} |r_i} = e \sum_{q}\sum_{q'}\beta_{f}^{q'\ast} \beta_{i}^{q} \braket{q'|\vec{r}|q}
\label{eq:melement}
\end{equation}
where $e$ is the electron charge and $\vec{r}$ is its position vector, 
while $q^{(\prime)}$ denote the full set of quantum numbers $n,l,j,m_j$.
Note that for an atom subject to only a magnetic or electric field, only 
a subset of basis states, determined by the selection rules, can be used 
for the diagonalization. But in combined electric and magnetic fields,
all the zero-field basis states with different quantum numbers $l$ and $m_j$ 
should be included, and already for small fields a set of more than 15000 basis 
states are required for the results of diagonalization to converge.

The precise polarization of the MW field with respect to the directions of 
the electric and magnetic fields at the excitation position is difficult to measure or simulate. 
Our calculations of the Rydberg state energies and transition dipoles as per 
Eqs.~(\ref{eq:epsr}) and (\ref{eq:melement}), assuming the magnetic field 
of $3.4\,$G along the $y$-direction, the electric field of $3.625 \,$V/cm 
in the $x$-direction and a MW field polarization in $z$-direction, well reproduce 
the observed transition frequencies, and we obtain the dipole moment 
$d \simeq 30 ea_0$ for the Rydberg transition $\ket{r_1} \to \ket{r_2}$. 
Changing the field strengths and directions can change the resulting transition dipole moment.\\

\textbf{Vacuum Rabi frequency.} 
The electric field energy in the cavity per MW photon is $CU_c^2=\hbar\omega_c$. 
With the cavity length $l=9.3\,$mm and the capacitance per unit length $C/l=164\,$pF/m, 
the (rms) ground state voltage on the central conductor is $U_c=3\,\mu$V. 
The resulting electric field per cavity photon at the position of the atoms $z = 130\,\mu$m below 
the chip surface and $0.85\,$mm away from the antinode is $E_c^{(0)}(z) = 0.015\,$mV/cm 
and varies by about 30\% over the extent of the cloud. 
With the MW polarization along the transition dipole moment, the vacuum Rabi frequency is then
$\Omega_{\mathrm{vac}} = d E_c^{(0)}/\hbar = 2\pi\times 0.6\,$kHz.\\

\textbf{Rydberg transition linewidth.}
The Stark gradients (static dipole moments) of the Rydberg states $\ket{r_1}$ and $\ket{r_2}$ 
are $d_{r_1} = -300\,$MHz/(V/cm) and $d_{r_2} = -463\,$MHz/(V/cm), leading to a differential Stark 
shift coefficient $\Delta d_r = d_{r_2} - d_{r_1} = - 163\,$MHz/(V/cm). As described above, a laser with linewidth
$\Delta \omega_{\mathrm{las}} = 2\pi\times 2\,$MHz excites the atoms to the Rydberg state $\ket{r_1}$ 
in a layer of width $\Delta z_{r_1,-}=0.28\,\mu$m, where the electric field varies by 
$\Delta z_{r_1,-} \alpha_{r_1,-} = \Delta \omega_{\mathrm{las}} /d_{r_1}$, as per Eqs.~(\ref{eq:Ergrad}) and (\ref{eq:Deltaz}).
Hence, the Rydberg transition $\ket{r_1} \to \ket{r_2}$ for the atoms in this layer is inhomogeneously
broadened by the spatially varying electric field by 
\begin{equation}
\Delta \omega_{12} \simeq \Delta z_{r_1,-} \alpha_{r_1,-} \Delta d_r = \Delta \omega_{\mathrm{las}} \frac{\Delta d_r}{d_{r_1}} \simeq  2\pi\times 1.1\, \mathrm{MHz} . 
\end{equation}

\subsection*{M4 IONIZATION SIGNAL}
\label{met:ion_signal}
The selective field ionization (SFI) signal for a specific Rydberg state and electric field ramp 
is calculated by following the time evolution of the atomic population through the Stark map, 
using the diagonalization of the Hamiltonian matrix with $2\,$ns step size. 
At each time step and for each state, an ionization rate is calculated using a complex absorbing potential, 
$\mathrm{CAP} = -i \eta W(\hat{r}, F_E)$, added to the Hamiltonian, with the scaling parameter $\eta = 1.52\cdot10^{-10} E_H/a_0$ (with Hartree energy $E_H$ and Bohr radius $a_0$), 
a $r^6$ radius dependence and a shift with the external electric field $F_E$ \cite{grimmel2017ionization}. 
The time dependence of the electric field corresponds to the voltage ramp at the extraction electrode, 
while the SFI signal has a time delay of $1.53\,\mu$s equal to the time of flight of the ions in the experiment. 

To limit the number of basis states and speed up calculations, the magnetic field is neglected during
the ionization process, avoiding thereby mixing the different $m_j$ states in the electric field.
This is justified as the interaction with the electric field is much stronger than the interactions 
with the magnetic field during the SFI field ramp. The SFI signal for different $m_j$ states can then be calculated separately.

The initial Rydberg states correspond to the Stark eigenstates in the combined magnetic and electric fields.
The state $\ket{r_1}$ has sizable contributions only from the $m_j=-5/2 \, ... \, 5/2$ substates 
of the $48D_{5/2}$ state, while the state $\ket{r_2}$ is given by a linear combination of zero-field states 
with different $l$ and $m_j$ values of the $n=47$ manifold. 
The resulting signal is then given by the sum over individually calculated 
(interpolated for all $l$ and $m_j$ values) ion signals weighted by the initial population distribution. 

\subsection*{M5 CAVITY DRIVEN RYDBERG TRANSITION}
\label{met:rabi_oscillation}

\textbf{Steady-state Rydberg spectrum.}
We model the interaction of an atom on the Rydberg transition $\ket{r_1} \to \ket{r_2}$ with the MW field 
by a driven two-level system. With the Rabi frequency $\Omega$, the decay rate $\Gamma$ 
and the resonant Rydberg transition frequency $\omega_{12}$, the steady state population 
of $\ket{r_2}$ is \cite{PLDP2007}
\begin{equation}
\rho_{22} = \frac{|\Omega|^2/4}{\left(\omega_{\mathrm{MW}}-\omega_{12}\right)^2 + \Gamma^2/4 + |\Omega|^2/2} . 
\label{eq:rho_22}
\end{equation}
To account for the Rydberg transition linewidth, we assume that the transition frequencies are distributed according to a Gaussian function
\begin{equation}
S(\omega_{12}) \propto \exp{ \left( -\ln 2 \left( \frac{\omega_{12}-\bar{\omega}_{12}}{\Delta\omega_{12}/2} \right)^2 \right) },
\end{equation}
where $\bar{\omega}_{12}$ is the mean transition frequency of the atoms in the resonant layer, and $\Delta\omega_{12}$ is the full width at half maximum. The appropriately weighted Rydberg state population is then
\begin{equation}
\tilde{\rho}_{22}= \int S(\omega_{12}) \rho_{22}(\omega_{12}) \, d\omega_{12}.
\end{equation}
The atoms interact with the intracavity MW field with Rabi frequency $\Omega = d E_c/\hbar$, where $E_c$ is 
given by Eq.~(\ref{eq:EcvsMW}). We denote by $\bar{\Omega} \equiv \Omega(\omega_{\mathrm{MW}} = \bar{\omega}_{12})$ 
the Rabi frequency of the injected MW field resonant with mean transition frequency $\bar{\omega}_{12}$ of the 
atoms in the resonant layer. We can then write
 \begin{equation}
|\Omega(\omega_{\mathrm{MW}})|^2 = |\bar{\Omega}|^2 
\frac{ (\bar{\omega}_{12}- \omega_c )^2 +  \kappa^2/4}{ (\omega_{\mathrm{MW}} - \omega_c )^2 + \kappa^2/4} . 
\label{eq:omegacavity}
\end{equation} 
The combination of Eqs. (\ref{eq:rho_22})-(\ref{eq:omegacavity}) then yields the model function 
$p_2 = \frac{p}{1+a} \tilde{\rho}_{22} + p_0$ for the normalized ion counts $p_2$ which we use to fit the data in 
Fig.~\ref{fig:4}\textbf{b} with the offset $p_0$, the amplitude $p/(1+a)$ and additional fitting parameters $\bar{\Omega}$, $\bar{\omega}_{12}$ and $\omega_c$. From the fit we deduce $\bar{\omega}_{12}=2\pi\times 20.5513\,$GHz, $\omega_c=2\pi\times 20.5495\,$GHz, and $p/(1+a)\approx 0.2$.\\

\textbf{Resonant Rabi oscillations.}
For a resonantly driven two-level atom, neglecting the population decay and coherence relaxation, 
the population difference between the states $\ket{r_1}$ and $\ket{r_2}$ is 
\begin{equation}
D(\Omega,t) = \rho_{11} - \rho_{22} = \cos(\Omega t)  .
\end{equation}
With the atomic cloud being laterally displaced by $x_0 \simeq 50\,\mu$m from the cavity center 
(see the inset in Fig.~\ref{fig:4}\textbf{c}), the MW field strength varies in the resonant atomic layer. 
We assume that the corresponding Rabi frequency varies approximately linearly in space, 
\begin{equation}
\Omega(x) = \bar{\Omega} \left( 1 + \frac{x-x_0}{\chi} \right) .
\end{equation}
The atomic density distribution in the layer is Gaussian, 
$n(x) \propto \frac{1}{\sqrt{2 \pi} \sigma} \exp{ \left( -\frac{(x-x_0)^2}{2\sigma^2} \right) }$. 
For the spatially averaged population difference we then obtain
\begin{equation}
\bar{D}(t) = \int_{-\infty}^{-\infty} n(x) D \big( \Omega(x),t \big) \, dx = e^{ -\gamma^2 t^2 } \cos(\bar{\Omega} t) 
\end{equation}
where $\gamma^2 = \frac{\sigma^2}{2\chi^2} \bar{\Omega}^2$. Hence, the spatial variation of the MW field
leads to damping of the resonant Rabi oscillations, even if we neglect the decay and coherence relations of the atoms. 

From the MW field simulations, we estimate the relative change of the Rabi frequency across the cloud of width 
$\sigma = 25\,\mu$m to be $b=\sigma /\chi \lesssim 0.3$, consistent with the observations in Fig.~\ref{fig:4}\textbf{c},
where we have  $\gamma_{\mathrm{fit}} =  1.6 \pm 0.2\, \mu\mathrm{s}^{-1}$, $2.7 \pm 0.6 \,\mu\mathrm{s}^{-1}$, and
$7.5 \pm 1.3 \, \mu\mathrm{s}^{-1}$ for $\Omega_{\mathrm{fit}} = 2\pi \times 1.7 \pm 0.1 \,$MHz,
$\Omega_{\mathrm{fit}} = 2\pi \times 4.3 \pm 0.2 \,$MHz and $\Omega_{\mathrm{fit}} = 2\pi \times 7.7 \pm 0.3 \,$MHz,
respectively. 
 
%%%%%%%%%%%%%%%%%%%%%%%%%%%%%%%%%

%%%%%%%%%%%%%%%%%%%%%%%%%%%%%%%%%

\section*{Acknowledgments}
This work was supported by the Deutsche Forschungsgemeinschaft through SPP 1929 (GiRyd), Project No. 394243350 (CIT) and Project  KL 930/16-1. 
C.G. acknowledges financial support from the Evangelische Studienwerk Villigst e.V.

\end{document}